\documentclass[aps,prl,twocolumn,groupedaddress,floatfix]{revtex4}

\usepackage{graphicx}  
\usepackage{dcolumn}   
\usepackage{bm}		
\usepackage{amssymb}
\begin{document}

\title{Mach cone in a shallow granular fluid}

\author{Patrick Heil}
\author{E. C. Rericha\footnote[1]{Electronic Address 
erin@chaos.utexas.edu}}
\author{Daniel I. Goldman}
\author{Harry L. Swinney\footnote[2]{Electronic Address 
swinney@chaos.utexas.edu}}

\affiliation{Center for Nonlinear Dynamics and Department of Physics, 
University of Texas, Austin, Texas 78712}

\date{\today}

\begin{abstract}

We study the V-shaped wake (Mach cone) formed by a cylindrical rod
moving through a thin, vertically vibrated granular layer.  The wake,
analogous to a shock (hydraulic jump) in shallow water, appears for
rod velocities $v_R$ greater than a critical velocity $c$.  We measure
the half-angle $\theta$ of the wake as a function of $v_{R}$ and
layer depth $h$.  The angle satisfies the Mach relation,
$\sin{\theta}=c/v_{R}$, where $c=\sqrt{gh}$, even for $h$ as small as
one particle diameter.
\end{abstract}

\maketitle

The interactions of a flow with an obstacle have  
been a test bed for fluid mechanics throughout the past 
century~\cite{tritton98}.  For example, measurements on the Von  
K\'{a}rm\'{a}n vortex street of flow past a cylinder, the drag force 
reduction due to turbulence, and shock wave interactions with airplanes 
continue to increase our understanding of fluid mechanics.  Experiments 
on granular flow past obstacles provide good geometries to test the 
emerging hydrodynamic 
theory~\cite{haff83,ahmadi83,lun87,jenkins85,goldshtein95,sela96,brey97a} 
for rapid granular flows.  In addition, studying the interaction of 
granular 
flows and obstacles is important for industrial applications.  Obstacles 
are often introduced to modify granular flows:
paddles are used to mix 
materials; inserts are added to granular bins to reduce stresses; and
pipes are inserted in chute flows as heat transfer surfaces~\cite{wassgren03}. 
  
Measurements on experimental and simulated flow fields indicate 
that shocks commonly develop when granular flows interact with 
obstacles~\cite{tan98,goldshtein95,buchholtz98,rericha02,bougie02}.  
Shocks form when the relative velocity between an obstacle and a fluid 
exceeds the wave speed in the medium.  The shock front is a  
superposition of waves excited as the  
obstacle moves through the fluid.   For a dispersionless Newtonian fluid 
with a 
constant wave speed, the front coalesces into a Mach cone with a half 
angle given by the Mach relation,
\begin{equation}\sin{\theta}={c\over{v_R}}\label{Mach},\end{equation} 
where $c$ is the wave speed and $v_R$ is the obstacle's velocity.  
Experiments and simulations on shocks in granular flow have not revealed an 
analogous relationship between $v_R$ and the shock angle.      

We study a phenomenon that is the analog of a compression shock: the wake 
that forms behind an obstacle moving on a free surface of a fluid
layer.  We measure the height field behind a thin  rod moving thorough 
a vertically vibrated granular layer.   We find the wake angle follows the 
Mach relation (\ref{Mach}) and is well described by the shallow
water theory for fluid flows without surface tension. In this theory, the
description of the wake is identical to that of a compressible shock in a
dispersionless gas.  Thus, the wake formed in the thin granular layer can 
be described by the usual tools of shock physics.

\begin{figure*}
\includegraphics[width=\linewidth]{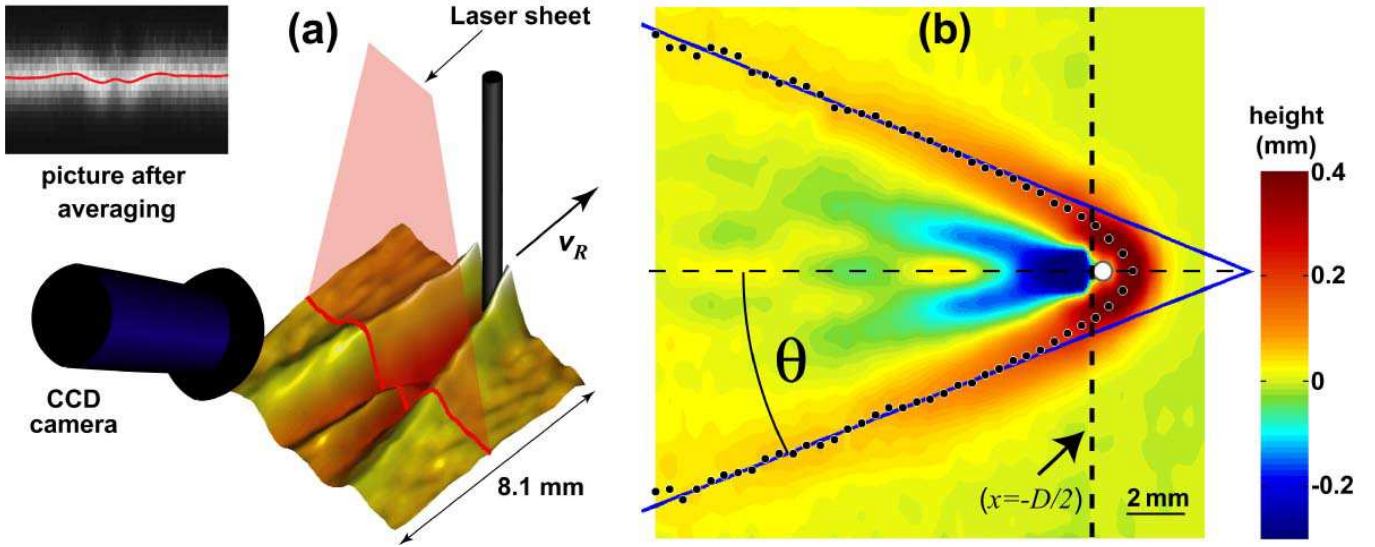}    
\caption{\label{fig:schematic}(color online)({\bf a}) Schematic of the
laser line scanning technique used to measure the displacement of the
surface created by a rod moving through a granular layer.  A laser
light sheet is incident downward onto the granular layer and is imaged
by a CCD camera set at a fixed angle with respect to the surface.  The
inset shows an image averaged over 400 periods of the rod motion.  The
location of the laser line is determined to subpixel accuracy by
finding the center of a Gaussian fit to each peak in the vertical
slice, as shown in the inset (red line).  ({\bf b}) Top view of the
shock created by the rod, moving to the right, for $v_R=21.5$ cm/s.
The location of the maximum layer height $\Delta h_{max}$ for each scan
line is indicated by the dots.  A linear fit to the maxima
(blue line) yields the wake's asymptotic half angle $\theta$.  A surface
height profile taken along the vertical dashed line is shown in the
inset of Fig.~\ref{fig:height}.}
\end{figure*}

{\it Experiment:} A stainless steel rod of diameter $D$=$0.75$ mm is 
inserted into a shallow, vibrofluidized granular layer consisting of
bronze spheres with a diameter $d=0.17$ mm.   The 
rod moves in a circular path of radius 51 mm with a constant speed $v_R$ 
in the range 4-30 cm/s.  

The granular layer is vibrofluidized using an apparatus similar to the
one described in~\cite{melo94,goldman04}.  For each layer depth $h$,
the peak plate acceleration 2.2$g$ and the
nondimensional frequency $f^*=f\sqrt{h/g}=$0.39 are chosen such that
the layer is fluidized but remains below the onset of
patterns~\cite{mujica98}.  The container is evacuated to less than 4
Pa to reduce air effects~\cite{pak95}.  The distance from the
bottom of the container to the rod is held fixed at 0.5$h$ throughout
the container oscillation.

We measure the time-averaged height field of the layer behind the rod
using a laserline technique similar to the one reported
in~\cite{forterre01}.  A thin vertical laser sheet (1 mm thick)
illuminates the granular layer.  When the rod passes through the laser
line it triggers a CCD camera, held at a fixed angle with respect to
the flat surface.  The camera captures 52 digital images of the laser
line separated in time by $\delta\tau=2.2$ ms
(Fig.~\ref{fig:schematic}).  For $v_R=21.5$ cm/s, the distance between
line scans is $v_R\delta\tau=0.47$ mm.  Deviations from a
straight laser line indicate the variations of surface height.  The
resulting height field is shown in Fig.~\ref{fig:schematic}.

The height field was averaged over many cycles with the frequency of
the driving and rod rotation incommensurate.  High speed
imaging showed that the angle of the shock front is independent of
phase during the driving cycle. The high speed images also showed that
the oval peak behind the rod (Fig.~\ref{fig:schematic}) is related
to the vibration of the layer during the cycle.

\begin{figure}
\includegraphics[width=\linewidth]{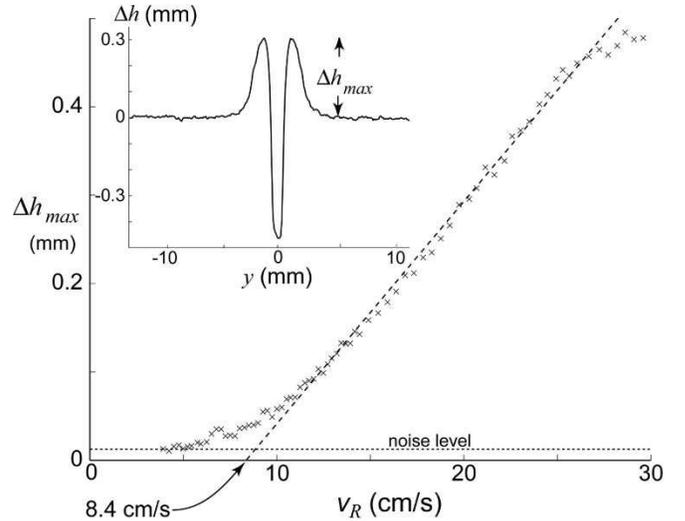}
\caption{The maximum upward displacement of the layer $\Delta h_{max}$
for the line scan at $x=-D/2$ for (inset shows profile for $v_R=23$
cm/s) as a function of $v_R$ for a layer depth of $h$=4$d$.  For small
$v_R$ the layer behind the rod remains flat to within experimental
accuracy (dotted line). Above a critical velocity the deflection
increases linearly with $v_R$(dashed line).  The intersection of the 
dashed line
with the horizontal axis indicates a nonzero critical velocity.  The
noise level was determined by the peak to peak oscillations in the
flat part of the layer.} \label{fig:height}
\end{figure}

{\it Results:} For small $v_R$, the time-averaged layer remains
everywhere flat to within our experimental error.  For $v_R$ greater
than a critical velocity $c$, the height field shows a bow shock
structure: a rapid increase in surface height, analogous to a
hydraulic jump, develops in front of the rod and extends downstream in
a V-shaped wake (Fig.~\ref{fig:schematic}).  The height profile taken
along the dashed line in Fig.~\ref{fig:schematic}(b) is shown in the
inset of Fig.~\ref{fig:height}.  The increase in height from the flat
layer to the maximum upward deflection of the layer ($\Delta h_{max}$)
measured from the laser line at $x=-D/2$ directly behind the rod
is shown in Fig.~\ref{fig:height}.  For $v_R>c$, $\Delta h_{max}$
increases linearly with rod velocity.  A fit to the data for a layer
depth of $h$=4$d$ indicates a critical wave speed $c=$8.4$\pm$0.7
cm/s~\footnote{Uncertainties in our experimental determination of the
critical velocity $c$ are due to the subjective choice of
the interval for the linear fits (dashed line in Fig.~\ref{fig:height}
and blue lines in Fig.~\ref{fig:schematic}(b).  Uncertainties in the layer 
depth $h$ are due to leveling of the container.}.

The transition from the subcritical flow without a shock, to a
supercritical flow with a bow shock is not sharp, as indicated by the
rounding of the transition seen in Fig.~\ref{fig:height}.  As the flow
accelerates around the rod, a small supercritical region develops for
$v_R$ less than but near $c$.  A shock forms in this region, but does
not extend out into the fluid.

\begin{figure} \includegraphics[width=\linewidth]{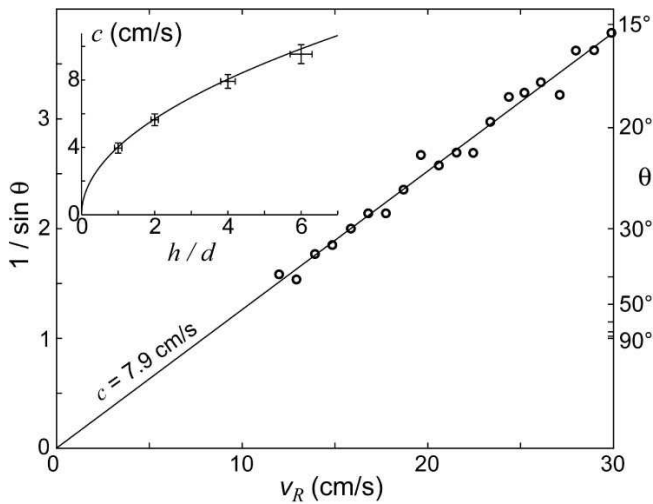}
\caption{\label{fig:sine} The measured dependence of the half angle
$\theta$ of the shock cone on $v_R$ is in accord with the Mach
relation (1) (solid line); the slope of the line yields
the wave speed $c$.  The inset shows the dependence of $c$ on the
layer depth $h$.  The curve is given by shallow water theory,
$c=\sqrt{gh}$.  The error includes uncertainty in the depth due to
leveling of the container (horizontal error bars) and uncertainty in
the algorithm determining the angle (vertical error bars).}
\end{figure}

We measure the half-angle $\theta$ of the shock with respect to the
axis of the rod's motion as a function of $v_R$.  We define the
location of the shock by the maximum of the height field for each line
scan.  Near the rod the shock is curved; however, within a few rod
diameters the shock straightens, creating a V-shaped wake with a
well-defined half-angle.  A linear fit through the maxima of the
asymptotic shock yields $\theta$ (Fig.~\ref{fig:schematic}).  We find
$\theta$ is described well by the Mach relation (\ref{Mach}) for a
compressible gas.  The linear dependence of the  data plotted in 
Fig.~\ref{fig:sine} indicates a constant surface wave speed.  For
$h=4d$ we find $c=7.9\pm 0.4$ cm/s, which is consistent with the
critical speed determined from the height measurement.

{\it Shallow water theory:} Our results can be understood in terms of
a shallow water approximation, similar to the approach applied to
avalanches~\cite{savage91,gray03} and granular free surface
patterns~\cite{bizon99b}.  When the depth of a fluid is small compared
to the other dimensions in the system, one can neglect the fluid
velocity in the vertical direction compared with the velocity
components parallel to the surface.  In this shallow water
approximation, the equations describing the motion of a free surface
of an incompressible, isothermal fluid in a gravitational field have the 
same form as the equations for a compressible gas
flow~\cite{landau&lifshitz}.  In both cases a shock forms when the
relative velocity between the fluid and the obstacle is greater than a
critical velocity.  For waves on a free surface the critical velocity
is the maximum gravitational wave speed, $c=\sqrt{gh}$, for long waves
without surface tension, and the shock is a discontinuity in height.
Our measurements for different layer depths yield surface wave speeds in
accord with the shallow water interpretation (inset to
Fig.~\ref{fig:sine}).

The agreement with shallow water theory for layer depths as small as
one particle diameter is surprising.  A vibrated granular layer is
highly compressible ~\cite{goldshtein95,bougie02}; during each
collision with the plate, a shock wave forms in the bulk of the fluid
and travels through the layer, compressing and heating the grains.
Although the volume fraction can change by a factor of two and the
granular temperature across the shock can increase by two orders of
magnitude~\cite{bougie02}, the energy is quickly dissipated by
inelastic collisions~\cite{bizon98,goldshtein03}.  Our molecular
dynamics simulations for the conditions of our experiment show that 
throughout much of the
cycle the bulk of the layer has an approximately constant density and
temperature.  The forcing of the plate acts only to fluidize the
granular layer and does not play a strong role in the propagation of
waves on the surface.  We expect the shock front in a shallow granular
layer to be unchanged under different methods for fluidization.

\begin{figure} \includegraphics[width=\linewidth]{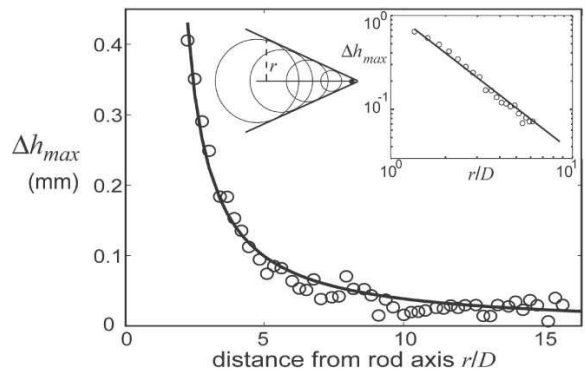}    
\caption{\label{fig:damping} The maximum height of the shock front as
a function of $r$, the perpendicular distance to the shock from the rod's 
axis of
motion.  The shock formed by the coalescence of cylindrical waves (inset) 
does not extend into the fluid indefinitely, but decays as $r^{-3/2}$ 
(solid
curve), as predicted by the Landau theory.  The line in the log-log plot 
(inset) has slope $-3/2$} \end{figure}

{\it Shock decay}: We find the granular shock rapidly decreases in
intensity as it propagates into the surrounding fluid.  The decay of
$\Delta h_{max}$ of the shock versus $r$, the distance from the shock
to the axis of motion of the rod, is plotted in
Fig.~\ref{fig:damping}.  The functional form of the decay agrees with
scaling predictions by Landau~\cite{landau45} for discontinuities in
cylindrical sound waves propagating into a dissipationless fluid.  The
velocity of each point in the shock front $u$ can be approximated by
$u=c_0+\left(\partial u/\partial\rho\right)_S \left(\rho_0 v_0/c_0\right) 
\sqrt{r_o/r}$, 
where $c_0$ is the wave speed of the
undistorted front, $\left(\partial u/\partial\rho\right)_S$ describes 
the adiabatic variation of wave speed with the local density, and 
$v_0\sqrt{r_0/r}$ accounts for the 
decrease in the intensity of a cylindrical wave as it
propagates away from its point of origin.  Conservation of mass 
requires the area of a shock profile remains constant as it moves.  As 
the intensity of the shock decays, the width of the shock increases.  
Setting the areas of shock profiles
separated by a time $\Delta t$ equal, Landau found that the width of
the shock increases as $r^{1/4}$, where $r$ is the distance from the
shock to the axis of motion of the rod (inset of Fig.~\ref{fig:damping}) , and 
the shock intensity $\Delta v$ decreases as $r^{-3/4}$.  For a shallow 
fluid, $\Delta v\propto\sqrt{\Delta h}$, implying $\Delta h$ should decay 
as $r^{-3/2}.$ The solid line in Fig.~\ref{fig:damping} is a fit to 
$\Delta h_{max}$ proportional to $r^{-3/2}$.

{\it Conclusions:} Our experiments demonstrate that a thin vertically
vibrated granular layer is described well by shallow water theory for
a surface-tensionless fluid.  We find that a shock forms on the
surface when an obstacle's velocity exceeds the speed of a gravity
wave, $c=\sqrt{gh}$.  The angle of the shock cone is determined by the
Mach relation, and the damping of the shock follows the scaling
derived by Landau for shocks traveling in a dissipationless fluid.
Future experiments should study the applicability of this model as a
function of layer depth and inelasticity.  For deeper layers, the
shock generated when the layer collides with the bottom plate may not
travel to top of the layer~\cite{goldshtein03}, possibly changing the
behavior.

The shocks formed in our experiment are an example of Cerenkov radiation
generated by an object traveling through a medium faster than the wave
phase velocity~\cite{jelley58}.  Such radiation leads to increased
resistance (wave drag) when a critical velocity is exceeded.  Future
experiments should examine the dependence of drag on $v_R$ near 
the onset of the
shock because experiments ~\cite{zik92,Burghelea01},
simulations~\cite{buchholtz98,wassgren03}, and 
theory~\cite{raphael96,chevy03} disagree on this increase in drag.
  
We thank Jon Bougie, Robert Deegan, W.D. McCormick, Larsson Omberg,
Jack Swift and Paul Umbanhowar for helpful discussions.  This research was 
supported by the Engineering Research Program of the Office of Basic 
Energy Science of the U.S. Department of Energy (Grant No. 
DE-FG03-93ER14312), by The Texas Advanced Research Program (Grant No. 
ARP-055-2001), and by the Office of Naval Research Quantum Optics 
Initiative (Grant No. N00014-03-1-0639). 


\end{document}